\documentclass{article}
\usepackage{authblk}

\usepackage{booktabs} 
\usepackage{amssymb}
\usepackage{amsmath}
\usepackage{array,multirow,graphicx}
\usepackage{lipsum}
\usepackage{subcaption}

\usepackage{amsthm}

\usepackage{balance}

\newtheorem{remark}{Remark}

\newtheorem{definition}{Definition}

\begin{document}
\title{Generation Meets Recommendation: Proposing Novel Items for Groups of Users\footnote{Pre-print. Accepted to Recsys'18.}}

\author{Thanh Vinh Vo}

\author{Harold Soh\thanks{Corresponding Author: harold@comp.nus.edu.sg}}

\affil{National University of Singapore}

\maketitle

\begin{abstract}
Consider a movie studio aiming to produce a set of \emph{new} movies for summer release: What types of movies it should produce? Who would the movies appeal to? How many movies should it make? Similar issues are encountered by a variety of organizations, e.g., mobile-phone manufacturers and online magazines, who have to create new (non-existent) items to satisfy groups of users with different preferences. In this paper, we present a joint problem formalization of these interrelated issues, and propose generative methods that address these questions simultaneously. Specifically, we leverage the latent space obtained by training a deep generative model---the Variational Autoencoder (VAE)---via a loss function that incorporates both rating performance and item reconstruction terms. We then apply a greedy search algorithm that utilizes this learned latent space to jointly obtain $K$ plausible new items, and user groups that would find the items appealing. An evaluation of our methods on a synthetic dataset indicates that our approach is able to generate novel items similar to highly-desirable unobserved items. As case studies on real-world data, we applied our method on the MART abstract art and Movielens Tag Genome dataset, which resulted in promising results: small and diverse sets of novel items.
\end{abstract}

\section{Introduction}
\label{sec:Introduction}

Automated recommenders~\cite{Resnick1997} have become a mainstay of modern information retrieval and e-commerce systems. By suggesting related items (e.g., news articles~\cite{Liu2010}, music~\cite{Celma2010}, movies~\cite{vig2012tag}), recommender systems filter away irrelevant data, enabling users to efficiently locate items of interest. However, current work has focussed primarily on providing recommendations of \emph{existing items} to consumers (either personalized to individuals or  groups~\cite{OConnor}). In this work, we take on the problem of recommending or proposing \emph{new} (\emph{non-existent}) items that could be \emph{created} to satisfy users. 

\begin{figure}
	\centering
	\includegraphics[width=0.8\textwidth]{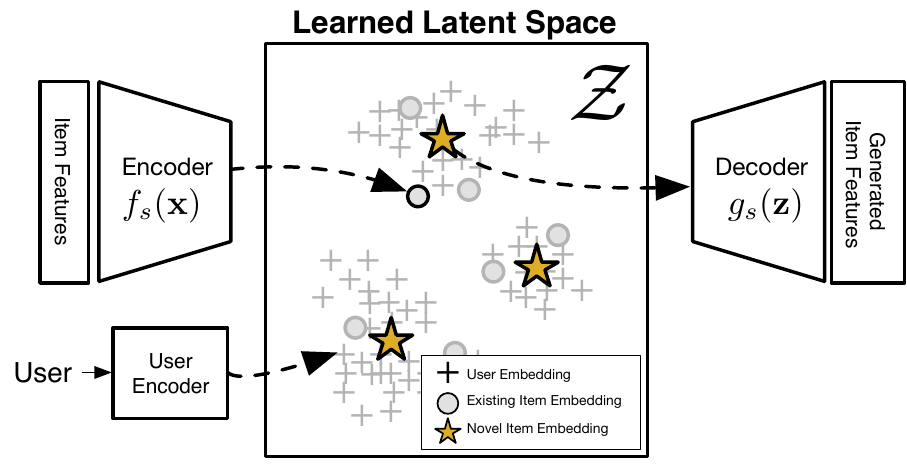}
	\caption{An Overview of Generating New (Non-Existent) Items for Groups of Users. Our approach relies on two related components: (i) a shared latent representation (embedding) space learnt from user-item rating data and (ii) a pair of functions: the encoder $f_s(\mathbf{x})$ that maps item features $\mathbf{x}$ to their respective latent representations/embeddings $\mathbf{z}$, and a decoder $g_s(\mathbf{z})$ that performs the inverse operation, i.e., maps embeddings $\mathbf{z}$ to item features $\mathbf{x}$. We exploit this space to obtain novel items with high predicted desirability through greedy weighted maximum coverage, and then generate item features via the decoder.}\label{fig:model}
\end{figure}

New item recommendation would be beneficial for item \emph{makers}---designers, producers, and manufacturers---who generate novel items for the marketplace. For example, mobile phone makers often produce a set of phone models because different users have different preferences; one group of users may prefer a phone with a large touch-screen, whilst another group may prefer a smaller, lighter, device. For item makers, important questions include: Is it possible to craft a set of new phone models that appeal to the widest possible number of customers? Which users would each phone design attract? And how large is each of these user groups?  These questions are currently \emph{not} answered by existing recommender systems research and may seem intractable at first glance.

In this paper, we take a first step towards a new class of recommender systems: item recommendation for item makers. Our goal is to develop a data-driven method for proposing new designs---potentially, by combining traits of existing items---that are predicted to be highly-rated. We propose a principled approach for addressing the aforementioned questions: first, we formalize the problem and show it to be a variant of maximum coverage, which is NP-hard in general. We then present an approximate scheme for generating plausible $K$ items that are predicted to be appealing to different groups of users. Our key insight is to leverage upon low-dimensional latent spaces (Fig. \ref{fig:model}), which we can efficiently search in a greedy manner (with a $1+\frac{1}{\epsilon}$ approximation guarantee). To learn this low-dimensional space, we propose two models: (i) a linear model similar to matrix factorizers~\cite{Koren2009a} and (ii) deep generative model, specifically, a collaborative variant of the Variational Autoencoder~\cite{kingma2013auto,Li2017} that is trained both to reconstruct ratings as well as observed item features. Both models enable projection of existing items into the latent space, as well as the generation of item features from ``imaginary'' item representations. 

We empirically evaluate our proposed scheme using a synthetic dataset comprising groups of users with different ``ideal'' items that are removed prior to model training. Results show that the deep model is able to recreate the missing ideal items with higher fidelity compared to the linear model. This results in better eventual coverage of the space. We also present results characterizing the sensitivity of the linear and deep models to the number of existing items, and describe the coverage of generated itemsets of varying sizes. As case-studies using real-world data, we used our methods to propose new art using the MART dataset~\cite{Yanulevskaya2012} and movies using the Movielens Tag Genome dataset~\cite{vig2012tag}. The set of resultant movie genomes is intuitive and diverse in nature; the top-3 generated movies are: (i) a sentimental/touching drama, (ii) a narrated social commentary with elements of dark-humor, and (iii) a big-budget action sci-fi film. 

To summarize, this work makes the following key contributions:
\begin{itemize}
	\item A formalization of the problem of recommending new items that can be created to satisfy different groups of users;
	\item A solution technique using greedy maximum coverage in a low-dimensional latent space learned via a deep generative model;
	\item An empirical evaluation of our techniques on  synthetic data and a demonstration of the approach on the real-world Movelens Tag Genome dataset.
\end{itemize}
From a broader perspective, this work takes the first crucial step towards intelligent item proposal for producers rather than consumers, e.g., we envision our proposed method can assist designers in the creative process. We discuss the limitations of the current approach and posit additional future directions in the concluding section of this paper. We hope this work will encourage others in the community to address this key problem, and look forward to a range of interesting solutions.

\section{Background and Related Work}
\label{sec:RelatedWork}

Research in recommender systems have flourished in recent years and we refer readers to survey articles~\cite{Resnick1997,Ekstrand2007,Koren2009a} for comprehensive reviews. Briefly, recommendation models have been proposed for a variety of items, from movies~\cite{Bell2007a} to tours~\cite{Anagnostopoulos2017}, and in a variety of contexts, e.g., with user and item features~\cite{Marlin2004,Stern2009,soh16}, 
and in sequential settings~\cite{Rendle2010,Soh2017a}. In this work, we bring together elements from group formation/recommendation and deep learning, which we briefly review below. 

\paragraph{Group Recommendation} Unlike standard individual user recommendation, group recommenders focus on suggesting items to \emph{user groups}~\cite{OConnor,Jameson07}, e.g., a tour~\cite{Anagnostopoulos2017} or a movie that all members consume together. Because individual users in the group may have different preferences, group recommenders attempt to balance the interests of all group members. This is typically achieved via two techniques: preference aggregation (e.g., \cite{Yu2006}) or score aggregation (e.g., \cite{Baltrunas2010}). In the former, recommendation is performed using a single prototype user obtained by combining the ratings of group members. In score aggregation, individual recommendations are combined via aggregation,  misery minimization, or game theoretic techniques~\cite{Amer-Yahia2009,Jameson07,OConnor,Zhang2017}. Recent work~\cite{Yuan2014} has also applied generative modeling, specifically a latent Dirichlet allocation (LDA), to better capture individual differences and influence. 

\paragraph{Group Formation} As a complementary problem to group recommendation, group formation aims to construct groups that would be maximally satisfied with an itemset \cite{Xiao,Mahyar2017,Roy2015}. Strategic recommendation aware group formation has been shown to be NP-hard under both least misery and aggregate voting criteria~\cite{Roy2015}. 

\paragraph{Deep Models for Recommendation} The success of deep learning~\cite{Deng2014,LeCun2015} in a variety of areas (e.g., computer vision~\cite{krizhevsky2012} and speech recognition~\cite{graves2013speech}) has inspired similar deep neural models for  recommendation~\cite{Karatzoglou17}. Examples include auto-encoders for collaborative filtering~\cite{Sedhain2015}, and recurrent neural networks for sequential recommendation~\cite{Soh2017a}. Very recent work~\cite{Li2017,Liang2018} have independently modified the Variational Autoencoder (VAE)~\cite{kingma2013auto} for recommendation, but with the purpose of modeling the underlying structure of observed data for existing items. 

\paragraph{Summary of Key Differences} Our paper differs from the above research in several notable ways: for example, the focus in group recommendation research is on trading-off the preferences within a given group. Here, the groups are not known in advance but emerge to comprise users that share an interest in a proposed novel item. Our latent space searching problem is also NP-hard, similar to group formation, but differs in formulation as we attempt to find novel item embeddings. Finally, unlike previously constructed collaborative VAEs which focus on making better item recommendations (e.g., under implicit feedback), we utilize the \emph{generative capabilities} of collaborative VAEs to \emph{construct} plausible new items; previous work had not considered nor evaluated this use-case.

\section{Problem Formulation}
\label{sec:ProblemFormulation}
In this section, we give a precise problem formalization for novel item generation for item makers; we first present an idealized maximization problem that is impractical to solve. We then make reasonable simplifying assumptions, leading to a practical variant that we address in this work. 

Given a set of users $\mathcal{U} = \{u_1, u_2, \dots, u_M\}$, we would like to find a set $K$ of \emph{novel} items $\mathcal{V} =  \{v_1, v_2, \dots, v_K\}$ that will be appealing to the users, i.e., that will maximize user ratings given by a rating function $r_{uv} = r(u,v)$. We assume that $\mathcal{V} \subseteq \mathcal{S}$, i.e., a subset of possible items $\mathcal{S}$. Furthermore, we want novel items that are highly-rated, above a given threshold $\tau$. We say an item \emph{covers} a user $u$ if $r_{uv} > \tau$. 
Given these elements, we would like to solve the following optimization problem which aggregates covered user ratings,
\begin{align}
	\text{argmax}_{\mathcal{V}} \sum_{u\in \mathcal{U}} \sum_{v\in \mathcal{V}} r(u,v) c_{uv}
	\label{eq:oriOpt}
\end{align}
subject to: 
\begin{align}
c_{uv} & \in \{0,1\} & \text{(if $c_{uv} =1$ then $u$ is covered by $v$)} \label{const:coverage}\\
\sum_{v \in \mathcal{V}} & c_{uv} = 1 &\text{(each user is covered by one item)} \label{const:nodoublecount} \\ 
|\mathcal{V}| & = K &\text{($K$ items are selected)} \label{const:maxksets}
\end{align}
Attempting to solve (\ref{eq:oriOpt}) above results in immediate practical challenges; the set of potential items $\mathcal{S}$ is unknown and we do not typically have access to the true rating function $r(u,s)$.

\subsection{A Latent-Space Formulation}
To make progress, we leverage upon latent real vector representation spaces. We assume that each item $s \in \mathcal{S}$ can be represented as a latent real vector (an embedding) $\mathbf{z}_s \in \mathcal{Z} \subseteq \mathbb{R}^{d_z}$. Although $\mathbf{z}_s$ is unobserved, each item is associated with a set of observed features $\mathbf{x}_s \in \mathcal{X} \subseteq \mathbb{R}^{d_x}$. For example, these features could be movie tags/descriptions or mobile phone properties (e.g., CPU, brand, screen size). 

Note that we are typically interested in obtaining a novel product's features $\mathbf{x}_v$ rather than its latent representation. As such, we assume that the item features $\mathbf{x}_v$ can be \emph{decoded} or \emph{generated} from $\mathbf{z}_v$ via a function $\mathbf{x}_v = g_s(\mathbf{z}_v)$. In other words, given $g_s$, finding an item's latent representation is equivalent to finding its features.

Likewise, we let each user be associated with a latent representation sharing the same space $\mathbf{z}_u \in \mathcal{Z}$. Users can also be associated with features $\mathbf{x}_u$ if available, but we do not focus on learning user features in this work. Crucially, the users and items are related by a pre-specified rating function $\hat{r}(\mathbf{z}_u,\mathbf{z}_s)$, e.g., the dot-product $\hat{r}(\mathbf{z}_u,\mathbf{z}_s) = \mathbf{z}_u^\top\mathbf{z}_s$. The key point is that since we do not know the true rating function $r$, we specify a convenient parametric form and learn the space $\mathcal{Z}$ that complies to observed ratings under $\hat{r}$.

\begin{definition}[MCNIP: The Maximum Coverage Novel Itemset Problem]
Given a specified rating function $\hat{r}(\mathbf{z}_u,\mathbf{z}_s)$, and a dataset $\mathcal{D} = \{ d^{(k)} = (u, i, \mathbf{x}_i, r_{ui})^{(k)} \}_{k=1}^{|\mathcal{D}|}$ of previously rated \emph{existing} items $\mathcal{I} \subseteq \mathcal{S}$ with associated features $\mathbf{x}_i$, we seek the latent representations $\mathbf{z}_v$, and the associated features, $\mathbf{x}_v$  for novel covering items $v \in \mathcal{V}$, where $\mathcal{V}$ is the solution to the following optimization problem:
\begin{align}
	\text{argmax}_{\mathcal{V}} \sum_{u\in \mathcal{U}} \sum_{v\in \mathcal{V}} \hat{r}(\mathbf{z}_u,\mathbf{z}_v) c_{uv}
	\label{eq:pracOpt}
\end{align}
retaining the same constraints (\ref{const:coverage}),  (\ref{const:nodoublecount}) and (\ref{const:maxksets}).
\end{definition}
We term this as the this \emph{Maximum Coverage Novel Itemset Problem} (MCNIP), due to its close relation to the well-studied maximum coverage problem (MCP)~\cite{nemhauser1978analysis}. 
As a by-product of solving MCNIP, we also obtain user groups, i.e., each group that is covered by an item $v$. MCNIP can be further generalized~\cite{cohen2008generalized} (e.g., to incorporate item costs and a total allowable budget) and we discuss possible extensions/variations in Section \ref{sec:Discussion}.

\section{Methodology}
\label{sec:Methods}
In this section, we detail our methods for solving MCNIP. At a high-level, our solution scheme consists of three basic steps:
\begin{enumerate}
	\item Learn the latent representation or embedding space $\mathcal{Z}$ along with the encoder $f_s(\mathbf{x})$ and decoder $g_s(\mathbf{z})$.
	\item (Approximately) Solve Eq. (\ref{eq:pracOpt}) to obtain a set of novel items $\mathbf{z}_v \in \mathcal{Z}_V$.
	\item Generate novel item features using the learned decoder $g_s(\mathbf{z}_v)$ for each $\mathbf{z}_v \in \mathcal{Z}_V$. 
\end{enumerate}
Step (3) is straightforward once $g_s(\mathbf{z}_s)$ is obtained, but steps (1) and (2) require additional elaboration. We describe both steps in detail in the following subsections. 

\subsection{Learning the Latent Space}
\label{subsec:LatentSpaceLearning}

Our goal is to learn $\mathcal{Z}$ given pre-specified rating function $\hat{r}(z_u,z_s)$ and a dataset $\mathcal{D}$ of user-item ratings. We also aim to learn an encoding function $f_s(\mathbf{x}; \phi): \mathcal{X}\rightarrow\mathcal{Z}$ and a decoding function $g_s(\mathbf{z}; \theta): \mathcal{Z}\rightarrow\mathcal{X}$. To reduce clutter, we drop the explicit dependence on the parameters when it is clear from context. The following descriptions assume dot-product-based latent spaces; in other words, the predicted ratings are a function $f_r$ of the inner product between the user and item latent representations, 
\begin{align}
\hat{r}(\mathbf{z}_u,\mathbf{z}_s) = f_r(\mathbf{z}_u^\top \mathbf{z}_s).	
\end{align}
but other reasonable rating functions can be easily substituted into the framework presented. 

In the following, we describe two models  where the encoder and decoder are either both linear or both nonlinear. Conceptually, we find $f_s$ and $g_s$ (or equivalently, the function parameters $\theta$ and $\phi$) by minimizing a total loss $\mathcal{L}$ that composed of three terms:
\begin{align}
	\mathcal{L} & = \text{Rating Loss} + \text{Reconstruction Loss} + \text{Regularizer} \nonumber\\
		& = \mathcal{L}_r(\theta) + \lambda_1\mathcal{L}_g(\phi) + \lambda_2\mathcal{R}(\theta, \phi)
	\label{eq:totalloss}
\end{align}
where $\lambda_1$ and $\lambda_2$ weigh the contribution of each loss component. Intuitively, the rating loss captures how useful the latent representations are for rating prediction; the reconstruction loss summarizes how well item features can be reconstructed from the latent vectors; and the last term applies regularization to avoid overfitting the dataset.

\subsection{Linear Encoding and Decoding} 
\label{subsec:linearmodel}
We begin with a straightforward model whereby both $f_s$ and $g_s$ are linear functions, 
\begin{align}
	\mathbf{z}_s & = f_s(\mathbf{x}_s) = \mathbf{M}_\theta\mathbf{x}_s \\
	\mathbf{x}_s & = g_s(\mathbf{z}_s) = \mathbf{M}_\phi\mathbf{z}_s
\end{align}
and $\mathbf{M}_\theta$, $\mathbf{M}_\phi$ are both real-valued matrices. When the rating function is $\hat{r}(\mathbf{z}_u,\mathbf{z}_s) = \mathbf{z}_u^\top \mathbf{z}_s$, this collaborative linear model (LM) is similar to hybrid matrix factorization (e.g., the Matchbox recommender~\cite{Stern2009}), but with an additional generative component.

The parameters $\theta = \mathbf{M}_\theta$ and $\phi = \mathbf{M}_\phi$ can be obtained by minimizing a loss of the form given in Eq. (\ref{eq:totalloss}), which we specify in turn. First, the choice of rating loss would depend on the the type of ratings observed. For example, if $\mathcal{D}$ contained binary ratings (e.g., ``likes'' or purchases), we could use the cross-entropy loss:
\begin{align}
	\mathcal{L}_r(\theta) = -\sum_{d \in \mathcal{D}}  r_{ui} \log(\tilde{r}_{ui}) +  (1-r_{ui})\log(1-\tilde{r}_{ui})
\end{align}
where $\tilde{r}_{ui}$ are the predicted ratings for observed items,
\begin{align}
	\hat{r}(\mathbf{z}_u,\mathbf{z}_i) = \sigma(\mathbf{z}_u^\top \mathbf{z}_i) = \sigma(\mathbf{z}_u^\top \mathbf{M}_\theta\mathbf{x}_i)
\end{align}
and $\sigma$ is the sigmoid function. For real-valued ratings (e.g., ``stars''), a squared error loss may be more appropriate. 
\begin{align}
		\mathcal{L}_r(\theta) = \sum_{d \in \mathcal{D}}  \| r_{ui} - \tilde{r}_{ui} \|_2^2
		\label{eq:ratingsqeuc}
\end{align}

Similarly, the type of observed item features would dictate the form of the reconstruction loss. For example, the cross-entropy loss would apply to item features that are binary or bounded in $[0,1]$ (e.g., grayscale image pixels):
\begin{align}
	\mathcal{L}_g(\phi) = -\sum_{d \in \mathcal{D}} \sum_j x_{ij} \log(\tilde{x}_{ij}) +  (1-x_{ij})\log(1-\tilde{x}_{ij})
	\label{eq:genxent}
\end{align}
where $x_{ij}$ and $\tilde{x}_{ij}$ are the $j$-th element of the observed $\mathbf{x}_i$ and reconstructed item feature $\tilde{\mathbf{x}}_i = \mathbf{M}_\phi \mathbf{z}_{i}$, respectively. 

As a specific example of combining these elements for a particular application, consider a model for generating Movielens tag genomes (with tag ``strengths'' in $[0,1]$) given a dataset of real-valued ratings; we can minimize the loss:
\begin{align}
	\mathcal{L}(\theta, \phi)  = &
	\sum_{d \in \mathcal{D}}  \Big[\| r_{ui} - \tilde{r}_{ui} \|_2^2\Big] - \lambda_1 
	 \Big[\sum_j x_{ij} \log(\tilde{x}_{ij}) + \nonumber \\
	 & \quad (1-x_{ij})\log(1-\tilde{x}_{ij}) \Big] + \lambda_2 \Big[\| \mathbf{M}_\theta \|_F + \| \mathbf{M}_\phi \|_F\Big]
	 \label{eq:linearmodelloss}
\end{align}
where we have combined (\ref{eq:ratingsqeuc}) and (\ref{eq:genxent}) with a regularizer comprising a  Frobenius norm for each matrix parameter, $\| \mathbf{M} \|_F = \sqrt{\sum_i\sum_j |M_{ij}|^2}$.


\subsection{Nonlinear Encoding and Decoding with Deep Generative Modeling} 
In this section, we derive a nonlinear model based on the variational autoencoder (VAE)~\cite{kingma2013auto}\footnote{We give a succinct derivation of the collaborative VAE and refer readers wanting more detail on the VAE and an alternative derivation to related work~\cite{Doersch2016,kingma2013auto,Li2017}.}. Although more technically involved than the linear version above, the principal ideas are not entirely dissimilar: the VAE also makes use of an encoder $f_s$ and decoder $g_s$, trained under a loss function to best reconstruct the items. A key difference is that these functions are now neural networks, which enable the model to capture more complex nonlinear relationships between the latent embedding $\mathbf{z}_s$ and the item features $\mathbf{x}_s$. 

The second key difference is that, unlike the deterministic linear model in the previous section, the VAE is derived under a probabilistic framework. As a guide, model development begins under the standard generative scheme where the objective is to find parameters $\theta$ that maximize the expected log probability of the dataset under the data distribution,
\begin{align}
	\text{argmax}_\theta \mathbb{E}_\text{data}[\log p_\theta(d)]
	\label{eq:maxlogpd}
\end{align}
where $d = (u, i, \mathbf{x}_i, r_{ui})$ is an observed datum with a user $u$, an item $i$ with features $\mathbf{x}_i$, and the rating $r_{ui}$. As we will see, the decoder is part of the specification of $\log p_\theta(d)$, and the encoder emerges as a means to perform (variational) inference. 

To begin, we specify the log-distribution within Eq. (\ref{eq:maxlogpd}),
\begin{align}
	\log p_\theta(d) = \log \mathbb{E}_{p(\mathbf{z}_i)}[p(r_{ui}|\mathbf{z}_u, \mathbf{z}_i)p_\theta(\mathbf{x}_i|g_s(\mathbf{z}_i))]
\end{align}
where $g_s(\mathbf{z}_i)$ is the decoder neural network with nonlinear layers. Unfortunately, computing $\log p_\theta(d)$ requires marginalizing out the latent variables $\mathbf{z}_i$\footnote{To simplify exposition, we focus on modeling the item embeddings as random variables and point-optimize the user embeddings $\mathbf{z}_u$; extending the model to treat $\mathbf{z}_u$ as random variables is straightforward given the SGVB sampling estimator.}, which is intractable in this setting. To get around this problem, we introduce the \emph{recognition} or \emph{inference} \emph{model}, $q_\phi(\mathbf{z}_i|\mathbf{x}_i)$ 
and perform approximate inference by maximizing the variational or evidence lower bound (ELBO), 
\begin{align}
	  \ell(\theta, \phi; d) \leq \log p_\theta(d).
\end{align}
where
\begin{align}
	\ell(\theta, \phi; d) = \mathbb{E}_{q_\phi(\mathbf{z}_i|\mathbf{x}_i)}&[\log p(r_{ui}|\mathbf{z}_u, \mathbf{z}_i)]  + \mathbb{E}_{q_\phi(\mathbf{z}_i|\mathbf{x}_i)}[\log p_\theta(\mathbf{x}_i|\mathbf{z}_i)] \nonumber \\
	&  -\mathbb{D}_\text{KL}[q_\phi(\mathbf{z}_i|\mathbf{x}_i)|| p(\mathbf{z}_i)]. 
	\label{eq:datumlb} 
\end{align} 
The above can be seen the expectation under the inference model with a KL divergence term that measures how different $q_\phi(\mathbf{z}_i|\mathbf{x}_i)$ from the prior $p(\mathbf{z}_i)$. Alternatively, it is the negative loss per datum; each of the three terms on the RHS correspond to the rating loss, reconstruction loss, and regularizer, in Eq. (\ref{eq:totalloss}).

In contrast to the linear deterministic model in Sec \ref{subsec:linearmodel}, each item is encoded as a probability distribution, $q_\phi(\mathbf{z}_i|\mathbf{x}_i)$, rather than a vector~\cite{soh16}. Here, we set $q_\phi$ to be a multivariate normal,
\begin{align}
	q_\phi(\mathbf{z}_i|\mathbf{x}_i) = \mathcal{N}(\mathbf{z}_i| \mu = f^\mu_s(\mathbf{x}_i), \Sigma = \text{diag}(\exp (f^\sigma_s(\mathbf{x}_i))) )
\end{align}
where $f^\mu_s$ and $f^\sigma_s$ give the mean and covariance of the distribution respectively, and the function $\text{diag}(\mathbf{b})$ returns a matrix with $\mathbf{b}$ along the diagonal. Thus, the pair $f^\mu_s$ and $f^\sigma_s$ forms the encoder.

Similar to the linear model, different rating and reconstruction likelihoods might be applied depending on the type of ratings and item features. As an example, for binary features, we could apply a Bernoulli likelihood:
\begin{align}
	\log p_\theta(\mathbf{x}_i|\mathbf{z}_{i}) = \sum x_{ij} \log(\tilde{x}_{ij}) +  (1-x_{ij})\log(1-\tilde{x}_{ij})
\end{align}
where $\tilde{\mathbf{x}}_i = g_s(\mathbf{z}_{i})$. The log-likelihood above has the same form as the cross-entropy loss given in (\ref{eq:genxent}). Applying a Normal likelihood results in a form identical to the squared error loss, up to a multiplicative factor and additional constant terms. The relationship between popular likelihoods and corresponding losses is well-known, and we omit further description here.

To train the model, we use the Stochastic Gradient Variational Bayes (SGVB) estimator, a differentiable unbiased estimator of the lower bound with smaller variance compared to direct Monte-Carlo estimation~\cite{kingma2013auto}. Using the SGVB amounts to using an approximate lower bound constructed from samples $\mathbf{z}^{(l)}_i$,
\begin{align}
	\tilde{\ell}(\theta, \phi; d) = \frac{1}{L}&\sum_l (\log p_\theta(\mathbf{x}_i|\mathbf{z}^{(l)}_{i}) + \log p(r_{ui}|\mathbf{z}_u, \mathbf{z}^{(l)}_{i})) + \nonumber \\
	& -\mathbb{D}_\text{KL}[q_\phi(\mathbf{z}_i|\mathbf{x}_i)|| p(\mathbf{z}_i)] 
	\label{eq:sampleloss}
\end{align}
where $\mathbf{z}^{(l)}_{i} = f^\mu_s(\mathbf{x}_i) + \exp (f^\sigma_s(\mathbf{x}_i)) \odot \epsilon_l$ and $\epsilon_l \sim \mathcal{N}(0, I)$. Using the auxiliary variable $\epsilon_l$ enables the random variable $\mathbf{z}_i$ to be specified as a deterministic function, and hence allows for differentiation through the sampling process~\cite{kingma2013auto}. This collaborative VAE model can be trained by minimizing the negative log-likelihood across the dataset:
\begin{align}
	\tilde{\mathcal{L}}(\theta, \phi) = -\sum_{d\in\mathcal{D}} \tilde{l}(\theta, \phi; d) 
	\label{eq:VAEtotalloss}
\end{align} 

To summarize, the inference network pair $f^\mu_s(\mathbf{x}_i)$ and $f^\sigma_s(\mathbf{x}_i)$ forms the encoder, and the network $g_s(\mathbf{z}_{i})$ is the decoder. Both are obtained by minimizing the loss function given by Eq. (\ref{eq:VAEtotalloss}) via using stochastic gradient descent. Compared to the linear model, the individual sample losses (\ref{eq:sampleloss}) have the same general form (\ref{eq:totalloss}), but the encoder/decoder are now neural networks embedded within a probabilistic generative framework.

\subsection{Latent Space Searching}
\label{subsec:LatentSpaceSearching}

Recall that we aim to find novel items that possess maximum coverage with maximum aggregate rating. Unfortunately, solving this problem exactly is impractical for large instances: 

\begin{remark} Given learned latent space $\mathcal{Z}$ and functions $g_s$ and $\hat{r}$, MCNIP with a finite discrete set $\hat{\mathcal{S}} \subseteq \mathcal{S}$ is NP-hard. 
\label{rem:nphard}
\end{remark}
\begin{proof}(\emph{Sketch}) The optimization problem (\ref{eq:pracOpt}) is exactly weighted MCP (a known NP-hard problem) with sets formed by the  covering sets induced by each candidate item $s \in \mathcal{S}$, and weights given by the rating function $\hat{r}$. 
\end{proof}

Fortunately, we can leverage upon a greedy algorithm for solving the weighted MCP problem, which achieves an approximation ratio of $1-\frac{1}{e}$ and is the best possible polynomial-time algorithm, subject to $P \neq NP$~\cite{nemhauser1978analysis}. The greedy algorithm is straightforward: successively pick the candidate item that attains the maximum total rating given by uncovered users until $K$ samples are selected. Note that if we are allowed to generate more then $K$ items, then we can achieve a better approximation ratio relative to the optimal $K$, e..g, if the greedy algorithm can pick $l > 5K$, the approximation ratio increases to $\approx 0.99$~\cite{nemhauser1978analysis,krause2014}.


To utilize the greedy approach, we first discretize $\mathcal{Z}$ to obtain candidate items. This can be achieved using several different techniques. With the collaborative VAE, a simple uninformed method is to sample new items from the prior $p(\mathbf{z}_s)$. Alternatively, if we assume that highly rated novel points will be similar to existing highly items, we can sample $T$ candidate points from $\prod^T_{t=1} q(\mathbf{z}_t | \mathbf{x}_t)$ where each $\mathbf{x}_t$ is selected from existing observed items $\mathcal{I}$ based on the rating, e.g., $p_r(\mathbf{x}_t| \mathcal{D}) = \text{Cat}(\boldsymbol{\alpha})$ where $\alpha_i = \frac{exp(\gamma\sum_u r_{ui})}{\sum_j \exp (\gamma\sum_u r_{uj})}$. 

As one might expect, higher levels of discretization (more samples) may yield better coverage but incur higher computational cost. In our experiments, we sample candidate items from a generative model trained using the posterior $\mathbf{z}_i$'s to ensure the sampling method was equal for the LM and VAE models. More precisely, we train a Gaussian Mixture Model (GMM) to learn $p(\mathbf{z}_s | \mathcal{Z}_\mathcal{I})$ where $\mathcal{Z}_\mathcal{I}$ is the latent representation of the observed items $\mathcal{I}$, and sample $T$ candidate items from  $p(\mathbf{z}_s | \mathcal{Z}_\mathcal{I})$.

\begin{figure}
	\centering
	\includegraphics[width=0.7\textwidth]{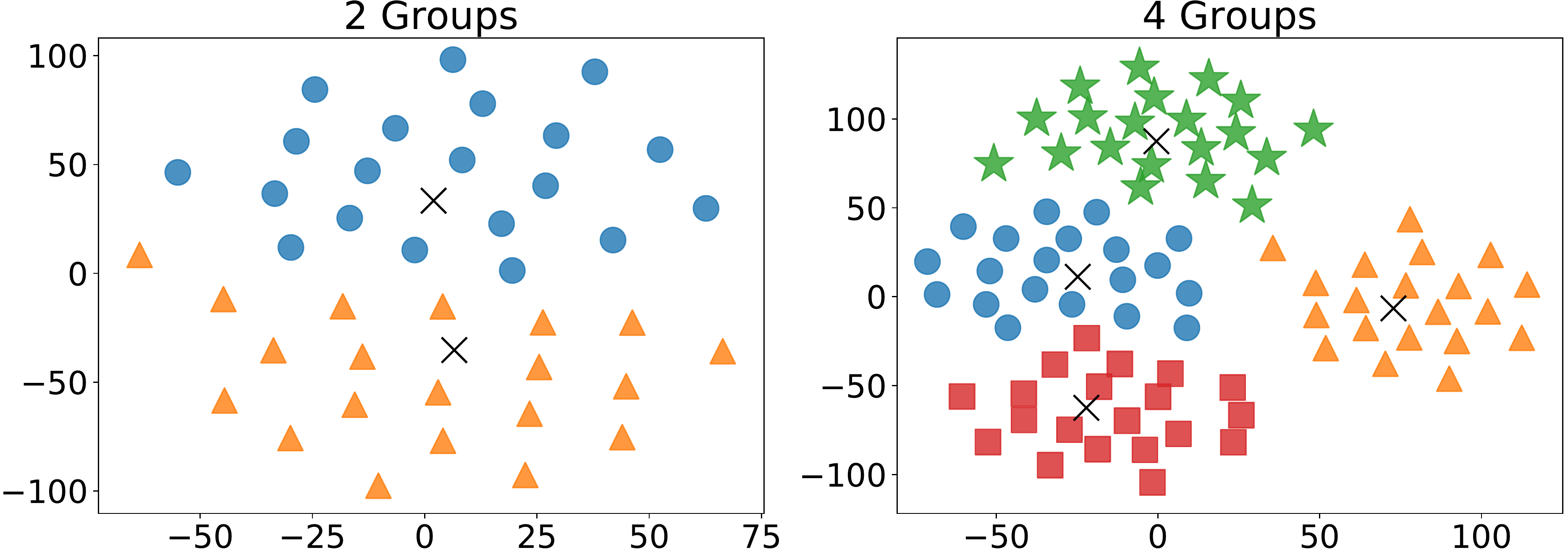}
	\caption{Groups and Observed 20-Dimensional Items in the Synthetic Dataset visualized in 2-D using t-SNE~\cite{maaten2008visualizing}. Ideal items (represented as $\times$'s) have the highest desirability but are unobserved. Our methods were tasked to recreate these ideal items.}\label{fig:syntheticitems}
\end{figure}

\section{Empirical Results}
\label{sec:Results}
Our primary experimental objective was to evaluate if our proposed approach is able to generate items that are both highly-rated and novel (different from existing items). As MCNIP is a new problem, there are currently no established protocols nor datasets to evaluate solutions. Thus, we crafted a synthetic dataset with \emph{unobserved} ``ideal'' (highly-desirable) items that we could compare our generated items against (described in Section \ref{subsec:Synthetic Data}). Next, to qualitatively evaluate our methods, we applied our approach to generate novel artwork and movie tag genomes~\cite{vig2012tag} as a case-study using real-world ratings and feature data (Sections \ref{subsec:art} and \ref{subsec:movielens}). 

\paragraph{Model Implementation and Training} We implemented the collaborative Linear Model (LM) and Variational Autoencoder (VAE) models in Tensorflow. All ratings data were rescaled to $[0,1]$ and we trained both models using the cross-entropy rating loss. We used the \textsc{Adam} optimizer~\cite{Kingma2015} with mini-batches of size 64, and learning rate of $10^{-3}$ with early stopping using a held-out validation set. 

\subsection{Generating Missing Ideal Items}
\label{subsec:Synthetic Data}

\begin{figure}
	\centering
	\includegraphics[width=0.6\textwidth]{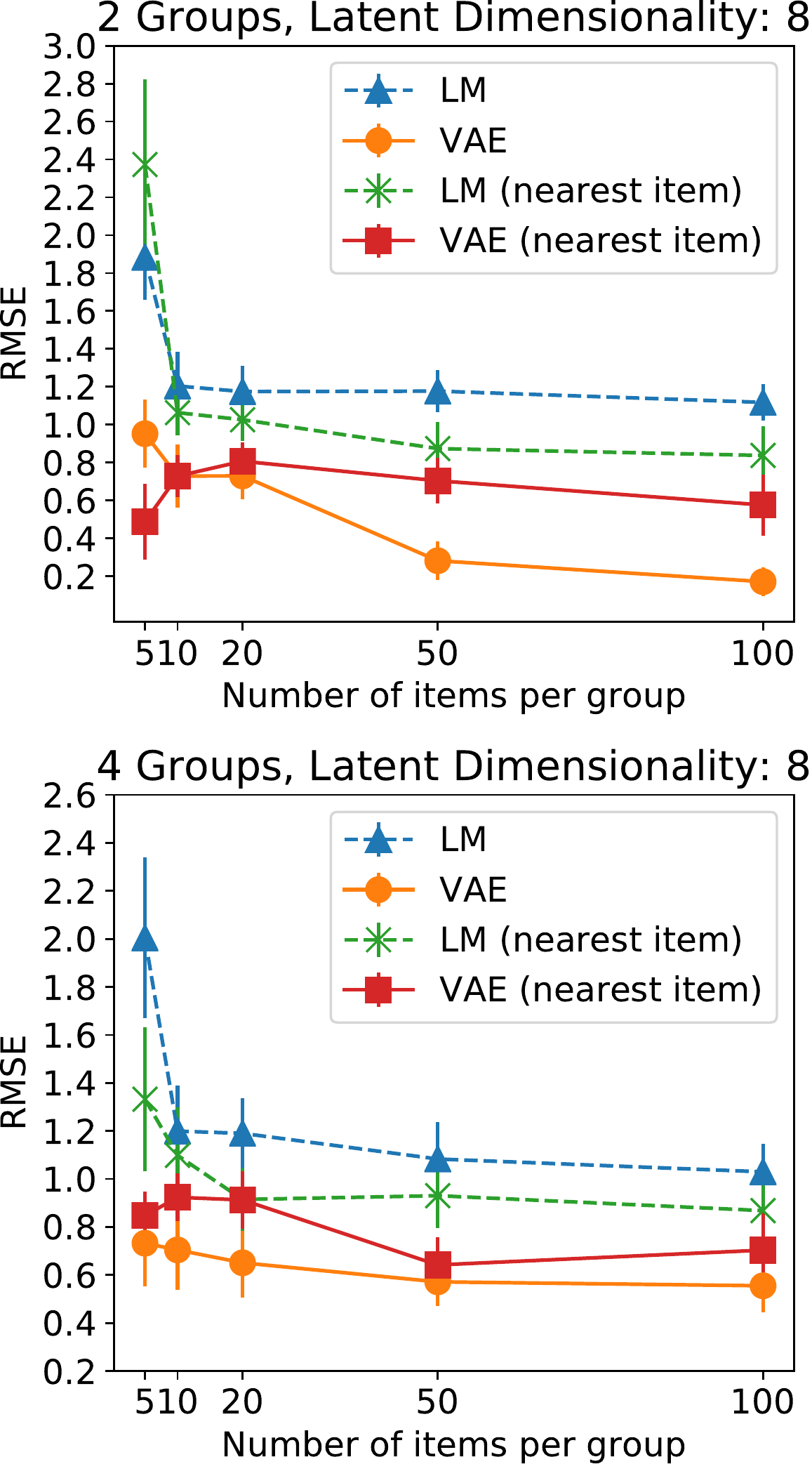}
	\caption{Novel item generation on synthetic data with two user groups (top) and 4 groups (bottom), and a varying number existing items. For both the collaborative Linear Model (LM) and Variational Autoencoder (VAE) models, the larger the number of existing items, the closer the generated item to the ideal item (as measured by RMSE). Across all the trials, the VAE was able to more accurately recreate the missing items compared to LM. The ``nearest item'' plots show the RMSE to the nearest existing item; the higher RMSE compared to the latent missing item indicates that the models were not simply generating existing items.}\label{fig:desireditems-synthetic}
\end{figure}


We constructed a synthetic dataset (Fig. \ref{fig:syntheticitems}) with 5000 users split evenly among $K$ groups. For each group, we sampled $N_g$ ``observed'' items $\mathbf{x}_i \in \mathbb{R}^{20}$ (20-dimensional feature vectors) from a multivariate Gaussian centered at the ideal item for that group, $\mathbf{x}_* \in \mathcal{X}_* \subset \mathbb{R}^{20}$. The observed items were rated differently depending on whether the user was from the same group or otherwise. For same group users, each item received a noisy rating inversely proportional to the distance to the ideal item $\mathbf{x}_*$. Items would obtain low ratings (0.1-0.3) from users belonging to other groups. Each user rated 20 items chosen randomly from the $K$ groups. 

To evaluate the sensitivity of our approach to the number of observed items and groups, we varied $N_g = 5, 10, 20, 50, 100$ items and $K=2,4$  to generate a variety of synthetic datasets. Note that the number of ratings was fixed across all the datasets since each of the 5000 users rated only 20 items. Using the observed items and ratings, we applied our methods---the underlying model was either the collaborative LM or VAE---to recreate the missing ideal items. The VAE  used neural networks comprising a single 14-neuron hidden layer with ReLU activations. Each method was repeated 12 times with different random initialization of their respective parameters. Our greedy algorithm used $2\times 10^5$ candidate points.

We measured performance by computing the RMSE of the generated covering items $\mathbf{x}_v$ to the unobserved ideal items $\mathbf{x}_*$; the matching of generated item to ideal item was performed by solving the standard linear assignment problem via a linear program. 

Fig. \ref{fig:desireditems-synthetic} summarizes our results, showing the VAE variant starkly outperforms LM across the different dataset parameters. The VAE is able to generate items that are significantly more similar to the missing ideal items. In addition, for the VAE, the RMSE between the generated item and the ideal item is less than any existing observed item, indicating that the VAE did not simply recreate a highly-rated existing item.
Fig. \ref{fig:desireditems-synthetic} also shows that both models generate better items when there are more observed items, potentially because a larger number of items enables the models to learn better encoding and decoding functions. 

\begin{figure}
	\centering
	\includegraphics[width=0.7\textwidth]{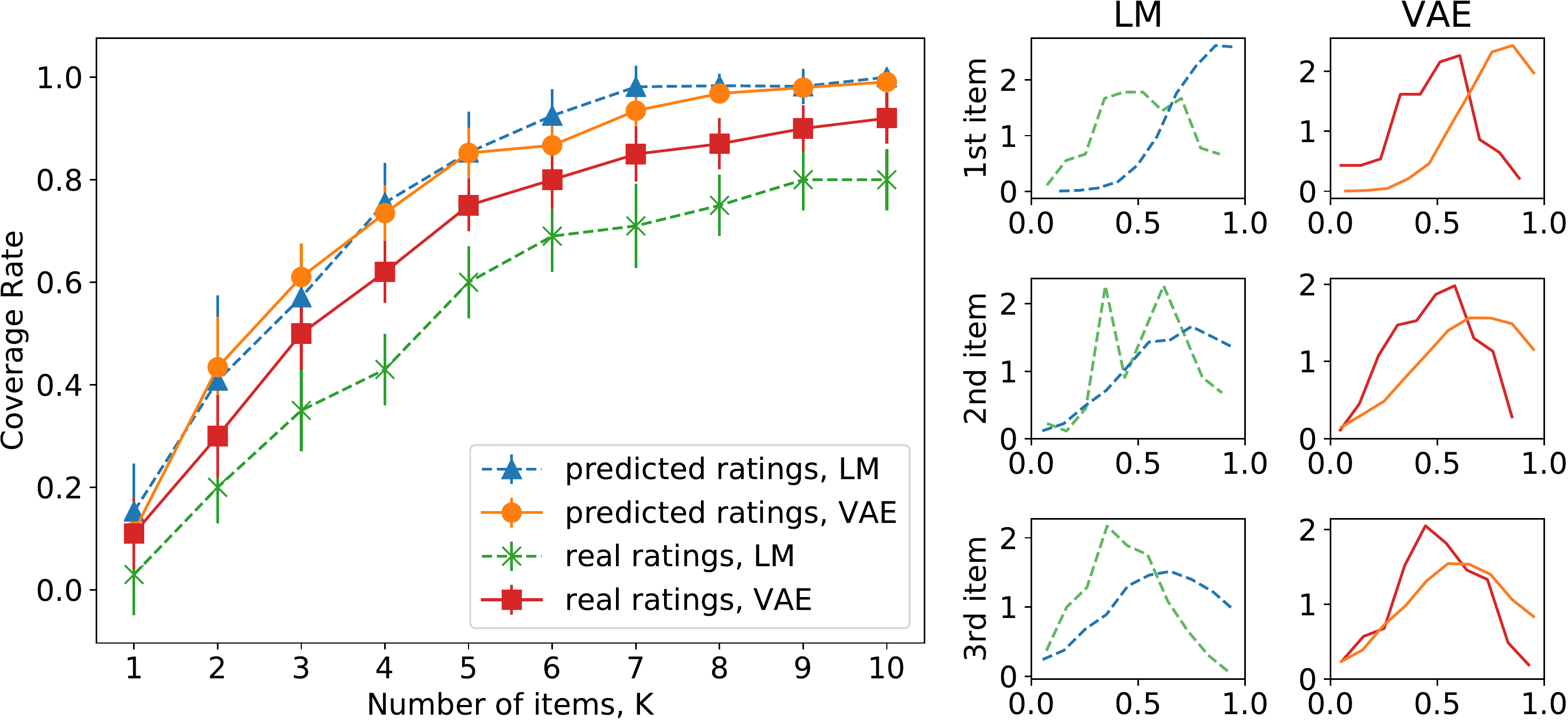}
	\caption{(Left) Coverage proportion on synthetic data (4 groups, 20 items per group) for the collaborative Linear Model (LM) and Variational Autoencoder (VAE). With higher $K$, a larger proportion of the users were covered ($\tau = 0.7$) with a diminishing rate of coverage. The predicted coverages were higher than the real coverages. (Right Subplots) The distributions of real and predicted ratings for both models show that the LM and VAE overestimated ratings (see left plot legend for line associations). This issue appears more problematic with the LM, and the VAE was able to attain higher real coverages.}\label{fig:coverage-synthetic-data}
\end{figure}

Fig. \ref{fig:coverage-synthetic-data} illustrates that the better generation also led to better coverage of the users (threshold $\tau = 0.7$), with VAE achieving a higher coverage than LM. As expected, with higher $K$, a larger proportion of users were covered, with diminishing returns at higher $K$. We also observed that the real coverage (computed using the real ratings), was lower than predicted. As the subplots in Fig. \ref{fig:coverage-synthetic-data} (right) show, both models tended to overestimate ratings. It may be possible to correct for this by tuning the loss component weights or using a different ratings loss function. Nevertheless, the results demonstrate that our generation technique using the VAE was able to generate items to satisfy the different groups of users.

\begin{figure}
	\centering
	\includegraphics[width=0.7\textwidth]{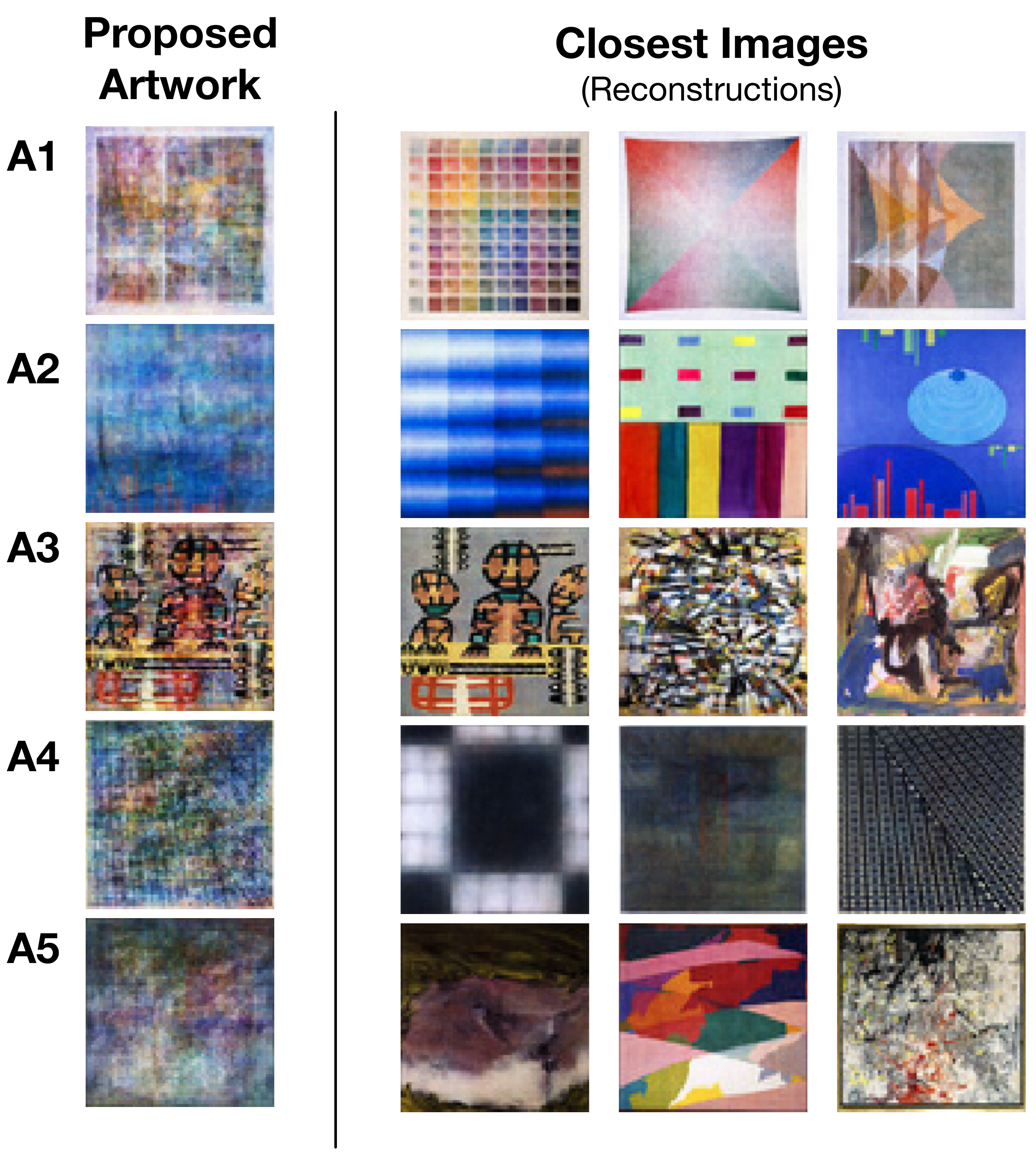}
	\caption{Generated Artwork using the MART dataset~\cite{Yanulevskaya2012}. Proposed artwork is shown in the left column, with the three closest artwork (reconstructed from the true images) on the right side.}\label{fig:genart}
\end{figure}

\subsection{Case Study 1: Generating Abstract Art}
In this first case study, we applied our methods to generating artwork. In this case, there are no ``ideal'' novel items for quantitative comparisons and as such, we evaluate our methods qualitatively. We used the MART dataset~\cite{Yanulevskaya2012}, which comprises 500 art pieces by professional artists and 10 thousand ratings by 100 people (74 females, ages 18--65). The ratings provided are positive emotion scores on a 7-point Likert scale, which we rescaled via decoupling normalization~\cite{jin2004}. The images were cropped and rescaled down to 64x64 pixels.   

We trained the collaborative VAE, which comprised three hidden ReLU layers of 512, 768, 384 and 256 neurons respectively, to learn a 128 dimensional latent state space. The method achieves a predictive accuracy of 62.5\% on a held-out test set containing 200 ratings.  We also attempted to train the linear model for comparisons but the generated images were poor (random looking pixels) and are not shown here. 

The generated artwork (with threshold $\tau = 0.9$ with $K=5$) are shown in Fig. \ref{fig:genart}. Although each proposed image possessed similarities to the existing artwork, they were also unique; we were unable to find exact matches in the dataset. Overall, the set tends to be colorful---possibly because the ratings are based on positive emotions---but is also visually diverse; A1--A5 have different textures and patterns. The entire set covers 82\% of the users, and the coverage of each painting A1 through A5 independently (the count of all users such that $r_{uv} > \tau$) were 61\%, 43\%, 44\%, 17\%, and 13\% respectively. Overall, we find these results promising; the generated art pieces suggest interesting novel combinations of forms and colors that may be appealing to users.


\begin{figure*}
	\centering
	\includegraphics[width=0.90\textwidth]{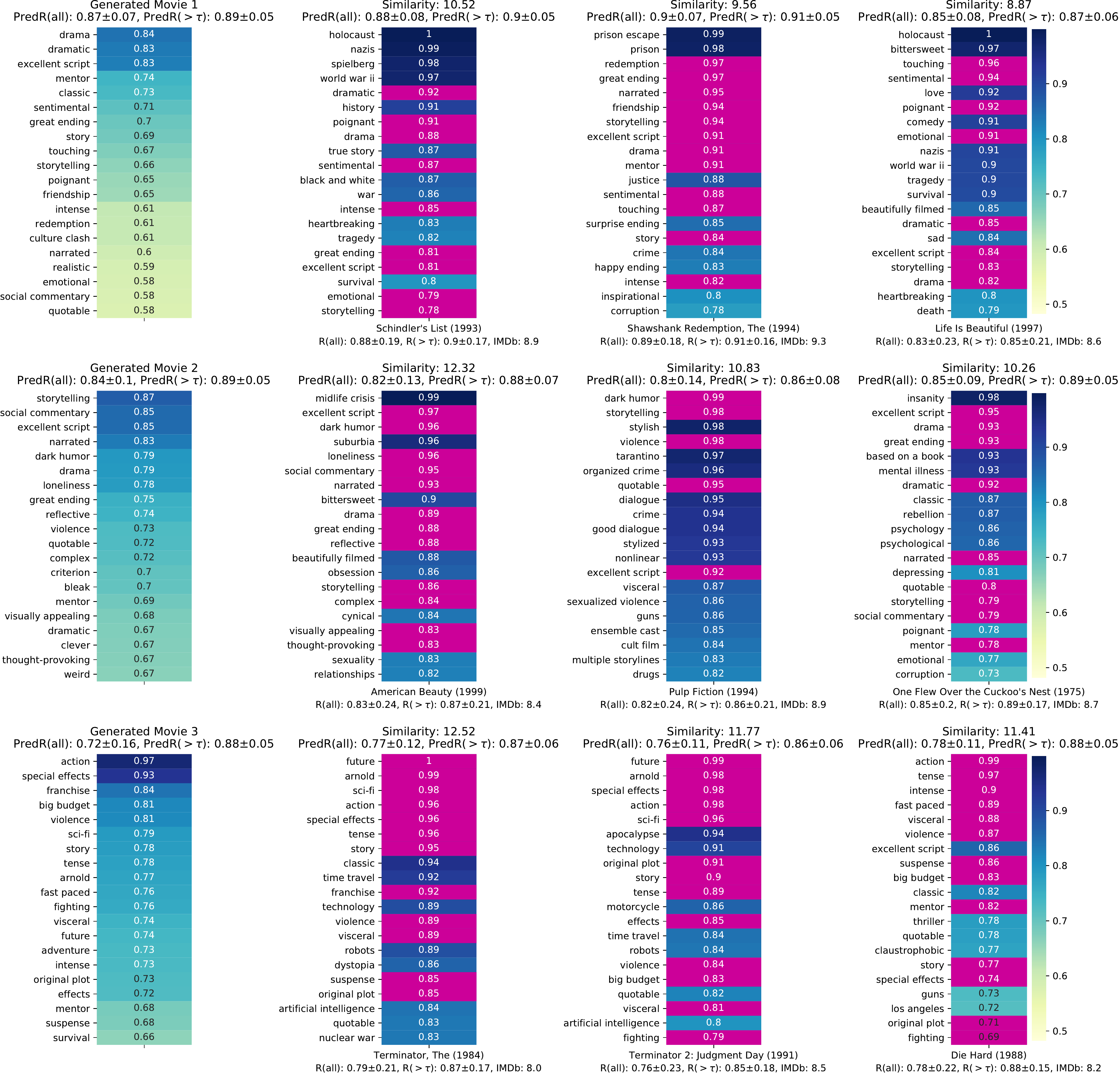}
	\caption{Top 20 Tags for the Generated Movie Tag Genomes (left column). The three most similar movies (as computed by the dot-product in the latent space) are shown in the same row as each generated movie. Pink rows indicate tags that are shared with the generated movie. PredR(all) and PredR($>\tau$) are the predicted ratings across all the users and users covered by the movie ($r_{ui} > \tau$) respectively. Likewise, R(all) and R($>\tau$) indicate the real average ratings. }\label{fig:generated-movies}
\end{figure*}

\subsection{Case Study 2: Generating Movielens Tag Genomes}
\label{subsec:movielens}
This section describes results obtained from applying our approach towards recommending novel movie features. We used the Movielens Tag Genome~\cite{vig2012tag} and MovieLens-1M datasets~\cite{Harper2015}, which contains 3952 movies, 6040 users and 1 million ratings. Each ``genome'' describes a given movie via 1128 tags with relevance scores between $[0,1]$. Here, we aim to generate new movie genomes, thus proposing new movies to create. Similar tags were merged (relevance scores were combined via a weighted mean using popularity scores) and non-informative tags (e.g., `good', `bad') were filtered away. After this preprocessing, each movie was represented by 450 relevance tags. 

We trained the VAE, which comprised three hidden ReLU layers of 500 neurons, to learn a 128 dimensional latent state space. To ascertain if the model learnt a reasonable representation for rating prediction, we compared the rating performance against matrix factorization (MF) and k-nearest neighbours (KNN). When applied to regular recommendation for existing items, the collaborative VAE achieves similar results to existing recommender systems (RMSE of $\approx$0.226). We then performed greedy maximum coverage with $K=5$ and $8\times 10^5$ candidate points. The linear model, while able to produce reasonable ratings, failed to generate interesting items and we focus on the VAE results.

Similar to the previous case-study, we evaluated our method qualitatively by comparing the generated movies against the most similar movies (in the latent space). The top-3 generated tags genomes (threshold $\tau = 0.8$) are shown in Fig. \ref{fig:generated-movies} along with similar real movies (as measured by the dot-product in the latent space). Interestingly, we find a diverse set of generated movies genomes: (GM-1) a sentimental/touching drama with a mentor and friendship; (GM-2) a narrated social commentary with elements of dark-humor and loneliness; (GM-3) a big-budget franchise action film with special effects and violence (possibly starring Arnold Schwarzenegger). The two other movies are (GM-4) a beautifully-filmed drama about culture-clash and childhood; (GM-5) an intense visually-appealing sci-fi movie with an emphasis on the future and technology. 

The cover for each movie GM-1 to GM-5 independently (the count of all users such that $r_{uv} > \tau$) were 84\%, 72\%, 36\%, 72\%, and 50\% respectively, suggesting all movies were predicted to be highly rated by large numbers of moviegoers. Each movie additionally covers a portion of the user base that was not previously covered. The predicted coverage of the set is 94\% and the top movie already attains a high predicted coverage of the users (84\%); all three closest movies to the top generated movie are critically-acclaimed movies that appealed to a large audience, and as such, the model predicted GM-1 to attract most users.


Although the generated items aren't finished movies, they illustrate promising ``genomes'' for new movies that would appeal to a wide audience.  Furthermore, they can be analyzed to determine important features: we observe certain repeated tags across all the generated movies---``excellent script'', ``mentor'', ``story''---that suggest common elements that highly-rated movies typically share.

\label{subsec:art}

\section{Discussion: Summary, Limitations, and Future Work}
\label{sec:Discussion}

The results shown in the previous section highlight the promise of the overall framework. As with any new venture, there are improvements that can be made to the solution scheme, as well as extensions to the problem formulation. The method is currently sensitive to the discretization procedure and the threshold parameter $\tau$; it is possible to generate a wide variety of interesting novel items by modifying these parameters. While this can be seen as a positive trait---a ``brainstorming'' process---it would be interesting to explore robust approaches. In the following, we highlight several other potential follow-up research: 

\paragraph{Novel Item Generation Under a Budget and other Constraints.} As mentioned in Section \ref{sec:ProblemFormulation}, the maximum coverage problem embedded within MCNIP can be extended. For example, we can consider the item maker to have a maximum budget $B$, and each candidate item $s$ to incur a cost $c_s$. This generalization can be accommodated in a straightforward manner by formulating the latent search as a generalized maximum coverage problem and using a modified greedy algorithm~\cite{cohen2008generalized}. A more difficult issue is to avoid proposing novel items that are infeasible, e.g., that violate physical law or current technological constraints. Potentially, this can be achieved by limiting the candidate set to only feasible items, but doing so in a computationally-efficient manner remains future work.

\paragraph{Complex Item Generation.} In this work, we have focussed on generating real vectors/matrices. However, real-world objects (e.g., artwork, mobile-phones, recipes) are considerably more complex. Generating such structured objects requires advances in generative modeling and we intend to investigate alternative models such as the variants of VAE (e.g., a hierarchical model~\cite{Zhao2017}) and alternative approaches such as the Generative Adversarial Network (GAN)~\cite{goodfellow2014generative}.
%
%

\paragraph{Collaborative Design with a Human-in-the-Loop.} Finally, it would be interesting to study how data-driven novel item recommendation can be used for  human-machine collaborative design. For example, a recent interactive photo manipulation tool~\cite{Zhu16} utilized a GAN to generate novel imagery based on human input. Similarly, we envisage that the methods proposed in this study can be utilized in a interactive design tool to help human item makers create desirable novel products. Research into interaction mechanisms and appropriate human-subject experiments could ascertain how such a tool could best aid the creative process.

\section{Conclusion}
\label{sec:Conclusion}

In this work, we have taken a first step towards solving the challenging problem of new (non-existent) item generation to satisfy groups of users. Our approach consisted of learning a latent real-vector embedding space using a deep generative model, and applying greedy weighted maximum coverage to locate promising items. Our experimental results on synthetic and real-world data indicate that good novel item features can indeed be produced from rating data. The advancement of these methods would lead to a new generation of recommender systems for item makers, rather than just consumers.

\balance
\bibliographystyle{ieeetr}
\bibliography{genrec}

\end{document}